# CDMA Technology for Intelligent Transportation Systems

Rabindranath Bera [1], Jitendranath Bera[2], Sanjib Sil [3],
Dipak Mondal [1], Sourav Dhar [1] & Debdatta Kandar [4]

[1]Sikkim Manipal Institute of Technology, Sikkim Manipal University,
Majitar, Rangpo, East Sikkim 737132, India
Phone: 03592-246220 ext 280; Fax: 03592-246112
E-mail : r.bera @rediffmail.com; r_bera@hotmail.com
[2]Department of Applied Physics, University of Calcutta
92 Acharya Prafulla Chandra Road, Kolkata 700 009, India
[3]Institute of Radio physics & Electronics, University of Calcutta
92 Acharya Prafulla Chandra Road, Kolkata 700 009, India.
[4]Department of Electronics & Telecommunication Engineering
Jadavpur University, Kolkata 700 032, India

*Abstract*

*Scientists and Technologists involved in the development of radar and remote sensing systems all over the world are now trying to involve themselves in saving of manpower in the form of developing a new application of their ideas in Intelligent Transport system( ITS). The world statistics shows that by incorporating such wireless radar system in the car would decrease the world road accident by 8-10% yearly. The wireless technology has to be chosen properly which is capable of tackling the severe interferences present in the open road. A combined digital technology like Spread spectrum along with diversity reception will help a lot in this regard. Accordingly, the choice is for FHSS based space diversity system which will utilize carrier frequency around 5.8 GHz ISM band with available bandwidth of 80 MHz and no license. For efficient design , the radio channel is characterized on which the design is based. Out of two available modes e.g. Communication and Radar modes, the radar mode is providing the conditional measurement of the range of the nearest car after authentication of the received code, thus ensuring the reliability and accuracy of measurement. To make the system operational in simultaneous mode, we have started the ' Software Defined Radio' approach for best speed and flexibility.*

## Introduction

Speed limit in the super highways is generally not imposed on the cars moving at their highest possible speeds. As a result, it often results in severe accidents and deaths. A CDMA radar based collision avoidance system can therefore be thought of which is to be fitted in the cars. This paper will highlight the detailed development of such radar for collision avoidance of cars.

CDMA Technology and its several versions are also popular for communication. It can also be exploited for a wide range of applications including range measurement, material penetration and low probability of interception. DS-CDMA, CDMA2000, MC-CDMA and MIMO CDMA [2],[3] are the different versions of same technology. The heart of such CDMA technology is the spread spectrum technology using PN sequence coding. The CDMA based digital radar technology will give rise to several advantages over conventional radars so that it can be used in ITS application successfully. Additionally, the same technology can also be explored to meet the communication need in ITS application [1].

## The Intelligent Mobile Campus Network (IMCN) [4][5]

The above mentioned two applications of CDMA in ITS can be further expanded in an IMCN which is modeled as shown in figure A. There are 4 cells namely cell1, cell2, cell3 and cell4 where each cell is defined as the geographical area of typically 100 meter over which a wireless communication is to be established between a mobile user and a fixed Base station. cell 1 and cell 2 are the two neighboring cells whereas cell3 and cell4 are another two remote neighboring cells. The four base stations will be placed on the rooftop of each building. All the four Base stations have their wireless connectivity with their respective wireless mobile handsets using the carrier frequency near 5.8GHz. The two neighboring base stations are connected by an MSC (Master Switching Center). So to have a total integrity among the four base stations two MSC namely MSC1 and MSC2 are required. As shown in figure A, MSC1 connects cell1 and cell2 whereas MSC2 connects cell3 and cell4. Each MSC is physically separated by a distance of 200 meter or more and is linked with the wireless network using 12 GHz microwave carrier. Thus the total system will provide full duplex communication with higher data rate of 64 Mbits/s approximately. Thus the total IMCN is too complex requiring a knowledge of several technologies for its design and successful implementation.

This paper will highlight only the CDMA technology and other relevant technologies used for the communication and radar applications utilized in the car. The authors are encouraged to exploit the latest digital communication and digital radar technology in their design and implementation.

### CDMA Technology in Car

A radio mounted on the car will normally face a lots of problems like:
1. Active Interferences comprising both adjacent as well as co-channel interference ( ACI & CCI).
2. Passive Interference coming from multi path .

### Multi path transmission

When the handset and base station are within line of sight, the primary propagation will usually be the line of sight and secondary propagations due to reflections will be less significant [6]. Reflected propagations become more significant if the line of sight is obstructed. Figure 1 illustrates a simplified multi path propagation. Whenever there is more than one significant impinging wave (with different phases) on a mobile receive antenna, the receiver will be subject to varying signal levels as it moves around. This is caused by constructive and destructive addition of the impinging waves due to their different phase offsets. This mechanism is called multi path fading.

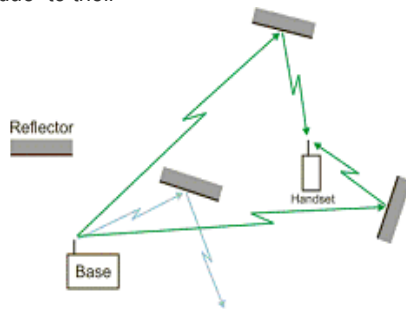

**Figure 1.** This figure illustrates a simplified multipath propagation, that will cause fading.

### Simulation of Space Diversity

The idea of antenna diversity is that if receive antenna A is experiencing a low signal level due to fading, also called a deep fade, Antenna B will probably not suffer from the same deep fade, provided the two antennas are displaced in position or in polarity. A Matlab based simulation is conducted in the Laboratory considering slow fading and the received signal strength variation is illustrated in figure 2.

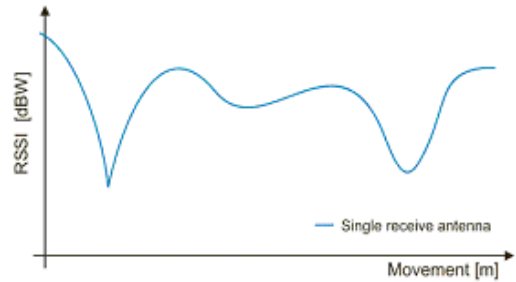

**Figure 2.** Typical Received Signal Strength Indication (RSSI) of a multipath faded signal.

The option to select the best antenna significantly improves performance in outdoor environments, but does not necessarily increase the maximum line-of-sight range of a product. Figure 3 illustrates the effect of selecting the best antenna. Antenna diversity is implemented by equipping the base station or handset (or both) with two antennas. Various selection schemes can be implemented, depending on the actual antenna setup. Preamble antenna diversity, also known as fast antenna diversity, has proven its use in fast frequency hopping systems. Preamble antenna diversity is implemented by comparing the RSSI value of each antenna in the beginning of each receive burst.

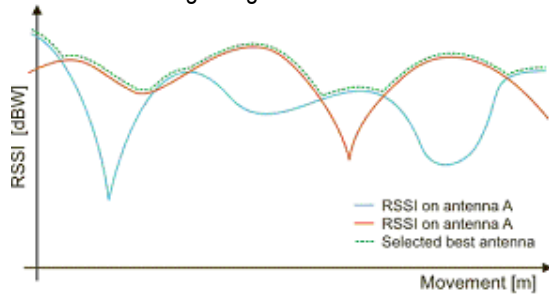

**Figure 3.** Demonstration of how two antennas can be used to ensure adequate signal level to the receiver.

### Experiments on Frequency Diversity

A delay spread multi path propagation study experiment was carried over the sea near Sagardwip Island ,The Bay of Bengal, West Bengal. A LOS link was set up over the sea saline water and two carriers at 12 & 13 GHz were transmitted simultaneously and received from a distance of about 5 km. The experiment was successfully conducted and the interesting results related to justification of using FHSS technology will be highlighted

The interesting observations are as follows:

a) A typical problem of fading with a fade depth of 30 dB is noticed at 12.5 GHz lasts for about half an hour depending on the sea water condition. The fading time varies from day to day. The LOS link data of signal strength at Kakdwip, received at Sagardwip over the river with saline water near Bay of Bengal is shown in fig.4.

b) With a deep interest to observe whether the above fading is effective at the same time to other neighboring frequency, we have transmitted two radio frequencies one at 12.5 GHz and another at 11.5 GHz and three kinds of observations are noticed and shown in Fig.4.

1: A typical problem of fading with a maximum fade depth of 30 dB is noticed at both the frequencies. It lasts for about half an hour depending on the sea water condition. The fading time varies from day to day.

2: The close observation of the results reveals that the signals of 11.5 & 12.5 GHz radio frequencies are not degraded at the same time, rather at different times.

3: The plots of the above observations are divided into 3 separate regions:

**Region I:** reception at both the frequencies is stable & normal

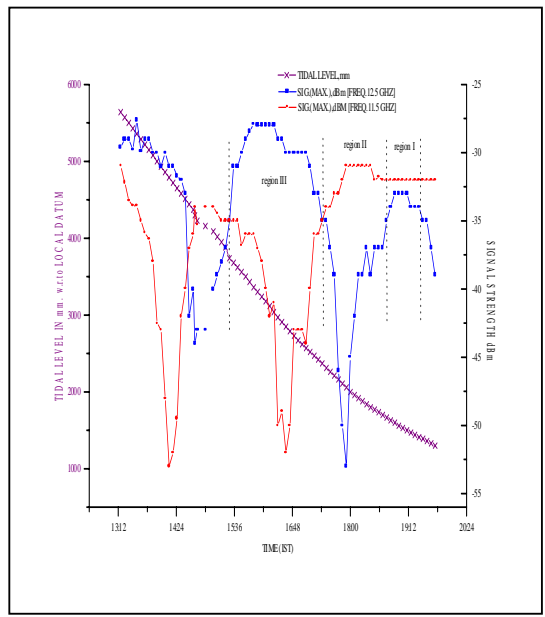

Figure 4. Frequency Diversity experiment

**Region II:** reception at 12.5 GHz is faded by approximately 30 dB while 11.5 GHz reception remains steady.

**Region III:** reception at 11.5 GHz is faded by approximately 30 dB while 12.5 GHz reception remains steady.

The reason for the above interesting observation is that the same path specified by sea water produces different time delay for two frequencies such that for one frequency signal strengths are additive and subtractive for other.

This complementary nature of fading at two different frequencies can be exploited to mitigate fading problem in a FREQUENCY DIVERSITY SYSTEM.

### Justification for Spread Spectrum (SS) Technology

In another experimental set up a SS based radio is tested for its Interference rejection capability. Figure5 illustrates the interference suppression capability of SS radio unit and is the right justification for the choice of SS in the physical layer. The region below the curve is the jamming free region.

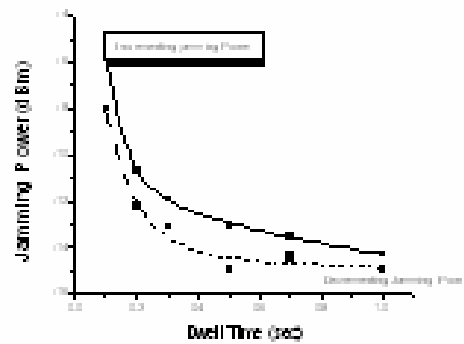

Figure5. Interference suppression capability of SS radio Vs. Frequency curve.

### Choice of Technology for ITS application

The above three Simulation and experimental facts can be referred to choose the proper technology for ITS application. Diversity reception may be the best choice to ( both frequency as well as space diversity ) restrict the active and passive interference to an acceptable level which are very strong in ITS application. Therefore, we like to incorporate the **FHSS** ( frequency hopping spread spectrum ) technology as a frequency diversion method and two antenna instead of single as **space diversity** reception system with a separation of 25 cm between them.

### Choice of frequency

To avoid the end user radio license problem, the best choice of radio carrier both for communication and radar should be the ISM band frequencies. There are 3 band of frequencies allotted for ISM bands, namely: 900 MHz, 2.4 GHz and 5.8 GHz.. The following Table 1 will dictate the choice of frequency.

**Table 1: Choice of Frequency for ITS**

| Freq. span (MHz) | Available bandwidth $\Delta f$ (MHz) for **frequency diversity** | Value of wavelength $\lambda$ (cm) | Effectiveness of **space diversity** | Remarks |
|---|---|---|---|---|
| 900-930 | 30 | 33 | Not effective | $\Delta f$ less, $\lambda$ more. |
| 2400-2480 | 80 | 12.5 | Effective | $\Delta f$ more, $\lambda$ less |
| **5760-5840** | 80 | 5.172 | **More effective** | **$\Delta f$ more, $\lambda$ least** |

The above table is self explanatory which will justify our choice of frequency for ITS application to be at 5.8 GHz.

**Specifications of the Radio**

| | |
|---|---|
| Transmission Type | **FHSS & Space Diversity** |
| Operating Frequency | **5.76-5.84 GHz** |
| Available Bandwidth | **80 MHz** |
| Separation between two antennas for space diversity reception | **25 cm** |
| Mode of Operation | **Communication ; Radar** |
| Interference Rejection | **Best** |
| Jamming resistance | **Good** |
| Delectability of nearby Car | **Good** |
| Voice quality | **Good** |

**Experimental Results**

Before launching any kind of radio application with the above specification (e.g. Communication or radar) the two channel characteristics, *bandwidth* and *Power*, constitute the primary resources available to the designer [7]. Accordingly, in an experimental laboratory set up (consists of developed radio and several Instrument like Pulse generator, Distortion analyzer, spectrum Analyzer, Digital Storage Oscilloscope etc.), the following results are established as shown in figure 6.

The radio is best usable in the range of 350- 600 micro second PRT range.

**Radar Modes of Operation**

A 13 bit code is transmitted from the Car using omni directional antenna so that signal from the nearby cars can be echoed back towards the space diversity reception system using two antenna. The space diversity antenna is helping us in determining the nearest car and rejecting reflection from distant car. The 'start' bit preceding the actual code is also helping in authentication of signal received i.e. if start bit followed by the received code then only the delay between transmitted and received code will be measured. It is then translated into the distance of the nearest car.

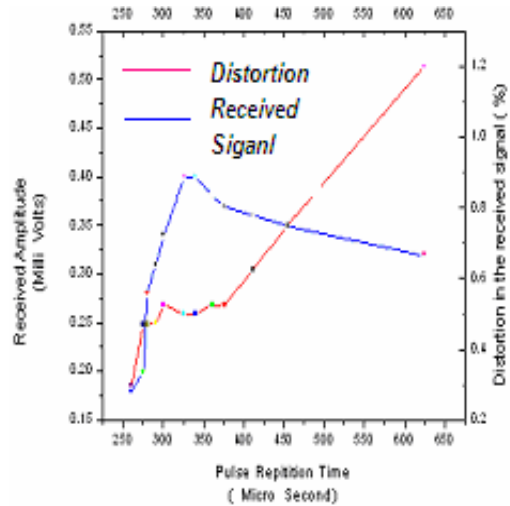

Figure 6; Channel Characterization of the Radio

**Conclusion**

The IMCN system is thus operational in its basic form for its two modes of operation supporting several mobile users over a distance of 500 meter or more. Lots of R&D efforts to be imparted for its commercialization.

**Further Extension**

The above radio model is successfully implemented at SMIT but is operational in 'either or' mode i.e. the same radio is either used as communication device or as radar installed in the Car. To make the system operational in simultaneous mode, we have started the ' Software Defined Radio' approach which is very much useful today for best speed and flexibility. The block diagram of revised ITS system is as shown in figure7. It utilizes two radio front end namely Radio I and Radio II serving the purposes of communication and radar respectively. At the backend, FPGA based signal processor will be used which will be finally controlled and monitored by PC. The other major components are A/D and D/A converter, Flash and SDRAM and 4 channel receivers.

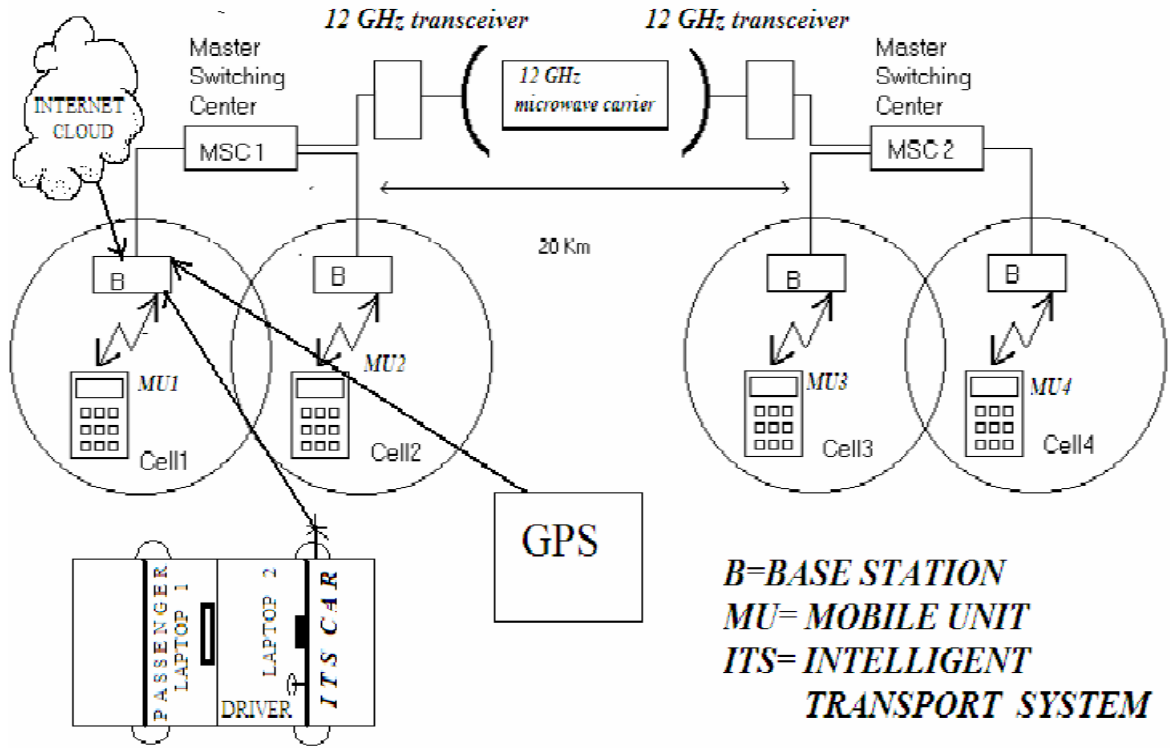

Figure A:  Intelligent   Mobile  Campus  Networking ( IMCN)  at Sikkim Manipal Institute of  Technology, Sikkim.

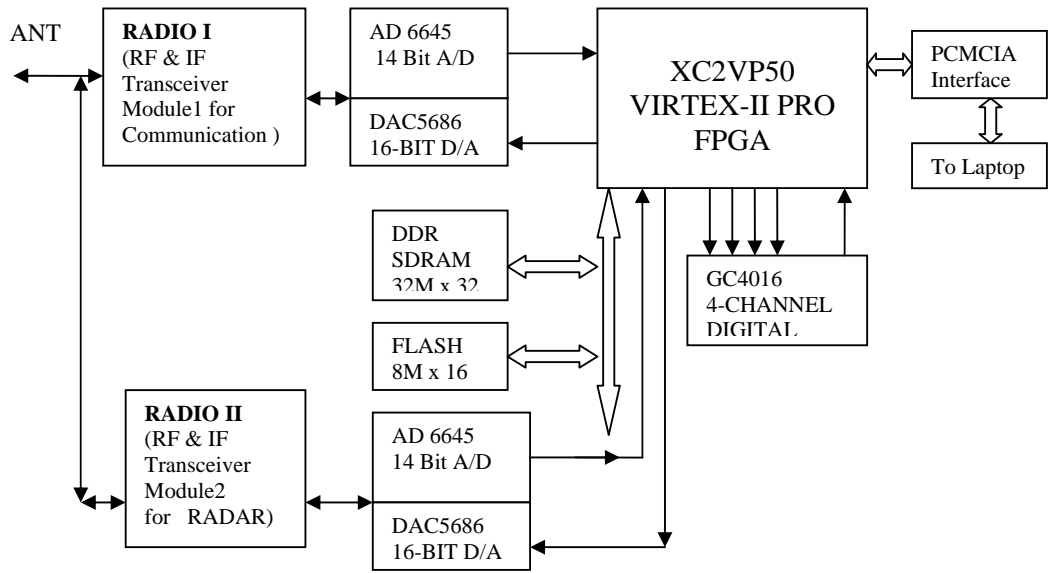

Figure 7: The block diagram of revised ITS system.

**Authors Information**

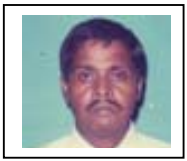
Rabindranath Bera: Born in 1958 at Kolaghat , West Bengal, B. Tech, M. Tech & Ph.D (Tech) from the Institute of Radiophysics & Electronics, The University of Calcutta, in the year 1982,1985 & 1997 respectively.  Currently working as Professor and Head of the Deparment, Electronics & Communication Engineering, Sikkim Manipal University, Sikkim, Microwave/ Millimeter wave based  Broadband Wireless Mobile Communication and Remote Sensing  are the area of specialisatoion.

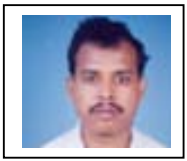
Jtendranath Bera: Born in 1969 at Sajinagachi, West Bengal, B. Tech, M. Tech from the Dept. of Applied Physics, The University of Calcutta, in the year 1993,1995 respectively, and Ph.D.( Engg.) from Jadavpur University, Kolkata. In the year 2005. Currently working as Reader, Dept. of Applied Physics, The University of Calcutta. Microwave/ Millimeter wave based  Broadband Wireless Mobile Communication, Remote Sensing and Embedded System  are the area of specialisatoion.

Sanjib Sil: Born in 1965 at Kolkata, West Bengal, B. Tech from IETE ( 1989), M. Tech from BIT, Meshra (1991). Ph.D. registration  from the Institute of Radiophysics & Electronics, The University of Calcutta ( 2002). Currently working as Asst. Professor, International Institute of Information Technology , Kolkata. Microwave/ Millimeter wave based  Broadband Wireless Mobile Communication, Remote Sensing  are the area of specialisatoion.

Dipak Mondal: Born in 1976 at Baruipur, West Bengal, B. Tech, M. Tech from the Institute of Radiophysics & Electronics, the University of Calcutta, in the year 2002, 2004 respectively. Currently working as Lecturer, Dept. of Electronics & Comm. Engg., Sikkim Manipal University, Microwave/ Millimeter wave based  Broadband Wireless Mobile Communication, Remote Sensing   are the area of specialisatoion.

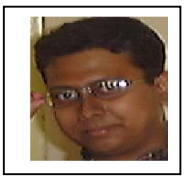
Sourav Dhar: Born in 1980 at Raiganj, West Bengal, B. E from Bangalore Institute of Technology, Visveswaraiah Technological University in the year 2002, M. Tech from Sikkim Manipal Instiutte Of Technology, Sikkim Manipal University in the year 2005. Currently working as Lecturer, Dept. of Electrical & Electronics, SMIT,  Broadband Wireless Mobile Communication  is  the area of specialisatoion.

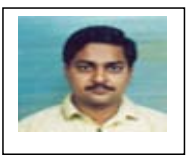
Debdatta Kandar: Born in 1977 at Deulia, West Bengal, B.Sc. ( Honours) from The University of Calcutta in the year 1997, M. Sc from Vidyasagar University in the year 2001. Currently working as  Research Fellow in Jadavpur University.